\documentstyle[12pt]{article}
\evensidemargin \oddsidemargin
\begin{document}
\pagestyle{empty}
\vspace*{1cm}
\begin{center}
\vspace{.5cm}
{\bf \LARGE Black Holes: A Window into } \\
\vspace{.5cm}
{\bf \LARGE A New Theory of Space Time} \\
\vspace*{1cm}
{\bf Spenta R. Wadia}{\footnote{\tt e-mail: wadia@nxth04.cern.ch, 

wadia@theory.tifr.res.in}}{\footnote{On leave from the Tata  

Institute of Fundamental Research, Homi Bhabha Road, Mumbai-400005,
India}} \\    

\vspace{.2cm}
Theoretical Physics Division, CERN \\
CH - 1211 Geneva 23, Switzerland \\
{\bf Abstract}
\end{center}
\begin{quote}
S. Chandrasekhar wrote in the prologue to his book on black holes,
"The black holes of nature are the most perfect macroscopic objects
there are in the universe: the only elements in their construction are
our concepts of space and time." In this contribution I briefly
discuss recent developments in fundamental theory and black holes that
vindicate this statement in a modern perspective. I also include some
of my reminiscences of Chandra.\\
\end{quote}
\vspace{.5cm}
\begin{quote}
\it{(Based on a talk given at the Commemorative Symposium on 

S. Chandrasekhar, at the Bhabha Atomic Research 

Centre, Trombay, India)}
\end{quote}
\vfill
\begin{flushleft}
CERN-TH/97-89\\
April 1997\\ 
gr-qc/9705001\\
\end{flushleft}
\vfill\eject
\pagestyle{plain}
\section{Reminiscences of S. Chandrasekhar}
My first reaction, when I was asked to give this talk to commemorate
Subrahmaniyan Chandrasekhar, was one of hesitation. Chandra, as he was
known to us at the University of Chicago, was a legendary figure
during his lifetime and one of the most distinguished astro-physicists
of the 20th century. His career spanned over 60 productive years and
given the volume, range and extraordinary scholarship of his work he
was an institution in himself. It is difficult to imagine the extent
of his achievement, especially within the exacting standards he set
for himself.  A lot has been written about Chandra and Kameshwar Wali
has written a very readable biography of Chandra. Here I would like to
add a few more strokes to the portrait of a man who was mentally a
very youthful and intense person even though he often compared himself
to Coleridge's 'Ancient Mariner'. I will take this opportunity to
narrate my first and my last meetings with him.

When I reached the University of Chicago in the fall of 1978, I wanted
to learn about Einstein's General Theory of Relativity because that
was a missing component of my education in physics. Fortunately
Chandra was offering a course on the subject during the semester and
that was an excellent opportunity. So I mentioned my interest to
him. Only later I found out that the classes were at 8.15 am in the
morning and almost impossible for me to attend. When I did not show up
for the first class he was very upset. Later that day he saw me as I
was riding in a crowded elevator. The elevator was so crowded that he
could not come in as the door opened. However in those brief moments
when the door remained open he saw me and pointing his umbrella at me
said, "You did not come because you are interested in fashionable
physics". The elevator door had closed by then and I walked to Eckart
hall for the weekly colloquium . During the colloquium tea (which is
held before the colloquium) I went up to him and tried to explain why
I could not come to his lecture and that annoyed him even more. He
walked away then reappeared and this is what he said, "You must have
heard of Johann Sebastian Bach. He used to wake up early every morning
and travel 16 miles to play the organ. But then neither you nor I are
Bach", and then he again walked away. That was my first encounter with
Chandra and my first glimpse of his sense of discipline and
humility. As time went by we developed a friendly relationship, but he
always looked upon me as a fashionable high energy physicist.

My last meeting with Chandra was in 1991 when I visited the University
of Chicago.  There was the usual appointment with his secretary, and
at the appointed hour he was waiting in his office. His mannerism had
a certain invariance about it. So did the arrangement in his office:
the portrait of Sir Isaac on the right hand wall, the bust of
Beethoven on the shelf behind. Books, volumes of the astrophysical
journal, manuscripts in calligraphic handwriting, were all in perfect
arrangement. There was a perfection and order in that office that I
have never seen anywhere else. He began by asking about how I was
doing and about the Tata institute and the Institute for Advanced
Study where I was spending the year. I wanted to be over with all this
as I was keen to show him a copy of a recent work I had done on a
subject that he was very interested in: black holes. He took the
paper, first turned to the references and asked, "Have you quoted my
book?" I said no. "Then why do you expect me to read your paper". I
was surprised at his response. I still wonder why it mattered so much
to him that we did not refer to his book. He then quickly flipped
through the pages and said, "There cannot be a black hole in 2-dims.".
When I told him about string theory and the dilaton that makes the
solution possible he responded, "You are throwing new words at me".

There was a silence afterwards and then he started speaking again,
softly and kindly reminiscing in a different vein, "You know Sir Isaac
never felt threatened by any of his contemporaries". The rest of what
he said I do not remember accurately, but it was to the effect that
Newton was such a towering genius that none of his contemporaries
bothered him.  He then mentioned how he has been working out the
proofs of the propositions in the Principia and how Newton's proof was
always better than the one he would construct.  I realised that
Chandra was focused on Newton during that time.

Since my visits to the University of Chicago were decreasing in
frequency, I had carried a camera with me to take some pictures of
Chandra.  I wanted to photograph him with the portrait of Newton in
the background. To my surprise Chandra took down the portrait and
placed it over the volumes of Newton's collected works that were on
his side table. Then he started hesitating to stand besides the
portrait of Newton! He felt so humble doing that. So I first took a
picture of Newton and then finally a picture of Chandra besides the
portrait of Newton.  An hour had gone by and I had to leave for
another engagement. That was the last time I saw Chandra.

When I went back to Princeton, I came accross V.I.Arnold's book,
'Newton and Hooke, Barrow and Huygens'. It was shocking to read about
Newton's feelings towards Hooke, to the extent that when Newton became
the President of the Royal Society, he destroyed every known portrait
of Hooke. I was bothered by this and wrote to Chandra. I never had
another opportunity to find out how he felt about this.

This last encounter with Chandra was a revelation for me and it
confirmed my feeling that Chandra had felt for many years that the
scientific community was not responding to him in the way he had
wished. That the changing fashions of science had left him
lonesome. There was little room for a man of his perfection. Besides
this, it also revealed to me that he had a fighting spirit till the
very end and that he too had his share of the common feelings that in
part shape the lives of most working scientists. It is this touch of
commonality that I find missing in the portraits of Chandrasekhar in
the literature.

\section{Black Holes and the Information Paradox}
In the following I would like to give a glimpse of some of the
developments in fundamental theory and the subject of black holes. I
had described the information paradox and related issues during my
talk but cannot resist from including more recent developments in this
contribution.

What are black holes?

Black holes are objects which divide space time into 2 regions which
are distinguished by a surface called the horizon. If there is a light
source inside the black hole then the light rays will be bent inwards
at the horizon and will never be seen from the outside. If one shines
light at the black hole from the outside it will seem to disappear at
the horizon and that is why the term black hole.  According to the
General Theory of Relativity black holes are likely to be the end
point of the gravitational collapse of massive stars and there is
mounting evidence that such objects do exist in the universe. There
are observations of black holes in binary systems and also as Active
Galactic Nuclei (AGN).

Besides being exotic objects which may hold a key to many unanswered
questions about the observed universe, black holes are of great
interest to theoretical physicists. When one quantizes matter that
interacts with a black hole according to the usual laws of quantum
mechanics one arrives at the conclusion, due to Hawking, that a black
hole is actually a black body whose temperature is of pure quantum
mechanical origin, $T={\hbar}{ \kappa}/2{\pi }$, where $\kappa$ is the
surface gravity of the black hole.  The black hole is in thermal
contact with the in-falling matter and there is a flux of Hawking
radiation characterised by the above temperature. Hence, in contrast
to the classical theory where in-falling matter disappears into a
black hole, the quantum mechanical wave function of the in-falling
matter extends 'accross' the horizon. Further, the black hole is
endowed with an entropy defined by the well known formula of Hawking
and Bekenstein, $S=A/{4G\hbar }$, where $A$ is the area of the horizon
of the black hole and $G$ is Newton's constant. The black hole behaves
like a macroscopic object obeying the laws of thermodynamics.

For a black hole of a few solar masses , the temperature ${T\sim
}10^{-8}$ degrees Kelvin and is obviously too small to be of any
practical significance. However black hole thermodynamics raises very
basic questions about the applications of quantum mechanics to black
holes.  Let me explain the issue. Since black holes can emit Hawking
radiation, they can evaporate by loosing their mass, and in case they
evaporate completely, the final state of the system would be simply
thermal radiation. Now this is problematic, because presumably the
black hole was formed by collapsing matter which was initially
described by a quantum mechanical wave function. Such a wave function
would describe many degrees of freedom, but the phase correlations
between the degrees of freedom would evolve in a unitary way,
according to the standard laws of quantum mechanics. In case the black
hole evaporates completely, and the final state is purely thermal
radiation then one is saying that the phase correlations are even in
principle completely lost. This is clearly a violation of unitarity as
we understand it in the standard formulation of quantum mechanics. It
is called the 'information paradox' of black hole physics.

Closely related to the above paradox is the question of the
statistical basis of the black hole entropy in accordance with
Boltzman's formula $S={kln(\Omega )}$, where $\Omega $ is the number
of microscopic states that constitute the black hole.

There is one more aspect of black holes that is quite disturbing. The
solutions that describe black holes have a curvature singularity,
where the standard laws of physics break down. The discussion of black
hole thermodynamics is not affected by the singularity because in most
cases the singularity is within the horizon (cosmic
censorship!). However the question of singularities has to be
addressed in a fundamental theory.

From the above it is quite clear that the application of quantum
mechanics to black holes reveals a possible inadequecy of either our
current notions of space time or of the standard laws of quantum
mechanics.  Many turning points in the history of science abound in
similar circumstances when two or more accepted theories lead to
contradictory answers for certain phenomena.  Both special relativity
and quantum mechanics grew out of the paradoxes that arose in the
applications of classical mechanics and classical electrodynamics to
phenomena like the propagation of light, black body radiation and the
stability of atoms. As we all know it was classical mechanics that was
replaced by quantum mechanics as the basic theory and classical
mechanics came to be understood as a limiting case of the former. The
measure of the deviation was provided by Planck's constant.

We see a similar situation in the application of quantum mechanics and
general relativity to black hole physics.  Hawking's view is that
quantum mechanics needs modification in the presence of black holes. A
different view has been advocated by 't Hooft, Susskind and many other
high energy physicists, who would like to retain the unitary
formulation of quantum mechanics and provide a standard statistical
explanation of Hawking radiation in an S-matrix framework.  Recent
developments seem to indicate that the unitary formulation of quantum
mechanics is retained and it is general relativity that is likely to
be replaced as a fundamental theory of space time. The theory that
replaces general relativity is called M-theory.  It a theory that is
presently being constructed and whose perturbative corners are
described by the 5 consistent superstring theories in 10-dim.  Of
course there is one essential difference between todays efforts and
the revolutions of special relativity and quantum mechanics in that
the latter were relatively quickly confirmed by experiment. Perhaps in
this respect M-theory may eventually have a historical parallel with
general relativity itself.  Even though M-theory is not in principle
inconsistent with any known experimental fact, there is presently no
prediction of M-theory that we can directly test experimentally. This
is due to the fact that the theory is not yet completely formulated
and also to the fact that perturbation theory does not seem adequate
to make contact with the phenomenology of elementary particles. This
difficulty of our situation makes the information paradox especially
important to tackle because it is more an issue of the logical
consistency of a framework rather than the agreement of predictions
with experiment. If M-theory can resolve the information paradox then
it would be very encouraging for our theoretical efforts. In what
follows we shall see that there is good evidence that M-theory is
indeed poised to resolve the information paradox.

\section{The Setting: M-theory}
M-theory resides in 11 space time dimsions, and its low energy limit
is described by 11 dim. supergravity theory. Its relation to the type
2A superstring theory in 10 dim. involves a compactification of the
11th dim.  to a circle of radius $r(\lambda )={\lambda ^{2/3}}$ ,
where $\lambda $ is the coupling constant of 2A theory. The type 2A
theory is one of the 5 consistent string theories in 10 dim. These
theories represent different corners of a large phase diagram which is
supposed to describe M-theory. One can device precise connections
between these string theories using duality transformations. An
essential role in this is played by certain elementary solitons of the
theory called D-branes. D-branes are domain walls of dimensions
ranging from $0,..,9$. A 0-brane is a point like object , a 1-brane is
a string like object, a 2-brane is a membrane and so on. These domain
walls are precisely defined by the fact that open strings end on them
and the end points of the open string can move freely inside the
D-brane. Hence the name D-brane, "D" stands for the Dirichlet boundary
conditions satisfied by the open strings in directions transverse to
the brane.  In this sense they are very much like the quarks in gauge
theories. They also carry integer units of charge (called the Ramond
charge) which can be measured by a tensor gauge field associated with
the brane. These solitons have a further remarkable property which
says that their mass and charge satisfy a precise relation,
$M=Q/{\lambda }$, and that this important relation is exactly true in
the quantum theory due to supersymmetry. One non-perturbative
formulation of M-theory, called M(atrix) theory in fact formulates the
fundamental theory entirely in terms of the open strings that connect
the simplest branes, namely the "0-branes". This is an expression of
reductionism taken to its extreme. I will not go into all these
exciting developments here but suffice it to say that in the setting
that we have sketched above it is indeed possible to give a
'constituent' model of a black hole which can give a standard
statistical basis to black hole thermodynamics in a unitary theory.

\section{D-brane Constituent Model of a Black Hole}
Let me now explain a specific constituent model, that describes the
5-dim.  black hole of type 2B string theory. This string theory is a
close relative of the 2A theory and is related to it by a duality
transformation. If one compactifies this 10-dim. theory on a
5-dim. torus, the long wavelength limit is described by a certain
supergravity theory, which has black hole solutions.  These solutions
are characterized by various integer charges. Among these the stable
black holes (those which satisfy the Bogomol'yni-Prasad-Sommerfield
(BPS) bound) are described by 3 positive integer charges. Let us call
them $Q_1,Q_5$ and $n$. $Q_1$ and $Q_5$ are the 'electric' and
'magnetic' charges corresponding to a tensor (2-form) gauge field. The
mass of this black hole is given by a linear combination of the
charges, and its entropy is given by $S = 2\pi \sqrt{Q_1 Q_5 n}$. This
stable black hole has zero Hawking temperature, but the non-zero
entropy signals a highly degenerate configuration. Small deviations
from this stable state are obtained by giving small values to other
charges that characterize the solution space.  These excited black
holes have a non-zero Hawking temperature and will certainly decay to
the stable black hole by the emission of Hawking radiation.  If one
scatters low energy particles from this black hole one can compute,
using the standard wave equation, the absorption cross-section and the
corresponding decay rate. The absorption cross-section at long wave
lengths turns out to be equal to the area of the horizon of the black
hole.

Now let me discuss the constituent model.  It turns out that the
constituents are an assembly of 1-branes and 5-branes.  There are
$Q_1$ 1-branes and $Q_5$ 5-branes.  The 5-branes are wrapped on the
5-torus and the 1-branes are wrapped on one of the circles of the
5-torus. To an observer in the 5 physical non-compact dimensions this
collection looks like a point. The collection of 1 and 5 branes
interacts via open strings between them, very much like the
interaction of quarks by gluon strings. Hence the low energy dynamics
is described by a $U(Q_1)\times U(Q_5)$ gauge theory.  This gauge
theory resides in the space time that is common to all the branes, and
in our example since the branes have only one dimension in common, the
gauge theory is 2-dimensional. It also turns out that this gauge
theory has N=4, supersymmetry. One of the implications of string
duality is that the same system can be described in terms of 0-branes
and 4-branes, which is described by a large $N$ matrix model.

We have a non-abelian gauge theory at the service of quantum gravity!

Before I proceed let me pause and explain the last statement.  To
illustrate the point consider the interaction between 2 D-branes,
separated by a certain distance $r$. This distance is measured in the
units of the "string length", which is a new length scale in the
problem. It is related to Newton's constant, which we would like to
take as a derived constant in the present framework. Since these
objects carry energy, their interaction will be mediated by gravitons
and other modes of the closed string. That is standard. As the
distance between the branes shrinks, the closed string description is
very complicated and involves the exchange of all the infinite number
of massive modes of the string.  However the propagation of a closed
string between the branes can also be visualized as a virtual open
string. While the long range interactions of the branes are best
described by gravitons, the short distance interactions between the
branes, in the limit of $r\rightarrow 0$, are best described by
massless open strings, which are the gauge fields. It is for this
reason that the bound state properties of the assembly of branes which
constitute the black hole is described by the infra-red properties of
the above mentioned non-abelian gauge theory. Another intriguing
implication of this picture is that since the small separations of the
branes are measured in terms of masses of the open strings that
stretch between them, the coordinates of the open string (gauge fields
in the transverse direction to the brane) are themselves space-time
coordinates. But these are matrices and in general do not commute!
Except when the branes are weakly interacting. This has led to the far
reaching suggestion of Witten that our usual notions of spacetime have
to be replaced, at distances smaller than the 'string' length, by a
non-commutative geometry. The gauge theory that describes the assembly
of D-branes actually embodies the space time degrees of freedom. This
idea finds its simplest expression when applied to an assembly of
0-branes as in M(atrix) theory.

Let me continue the discussion of the black hole bound state. To
describe a macroscopic black hole we clearly need a large number of
these branes and in fact we have to choose $Q_1$ and $Q_5$ to be
comparable and large. Let us call that large number $N$. A sensible
limit is one in which $gN$ is held fixed as $N{\rightarrow \infty}$
and $g{\rightarrow 0}$. Here $g$ is the coupling constant of the
string theory. The gauge theory then has a systematic expansion in
powers of $1/N$.  The analogy with $SU(N)$ QCD in the large $N$ limit
is instructive. As is well known the low energy effective lagrangian
of QCD is the chiral model of mesons whose expansion parameter is
given by the inverse of the pion coupling constant ${f_\pi}\sim
1/N$. The baryon is an $N$ quark bound state interacting via gluons
and it is a soliton solution of the chiral lagrangian with $M\sim
N$. The baryon is analogous to the black hole in that it is composed
of $N^2$ open string degrees of freedom and it is also a classical
solution of a low energy supergravity lagrangian. Newton's coupling
scales as $G\sim 1/N^2$. The mass of the black hole scales as $M\sim
N^2$.  Just like in QCD, where meson-baryon couplings are of order
$(1/N)^0$ and of the same order as the pion kinetic energy term, the
closed string- black hole couplings are also of order $(1/N^2)^0$, of
the same order as the graviton kinetic energy term. In both cases the
interaction is of order one and that is why there is a non-trivial
scattering. Also, the size of the baryon is independent of $N$ and so
is the area of the horizon of the black hole. However the analogy is
partial because the lowest lying collective modes of the baryon are
described by the collective coordinates of the flavour group, and
hence the degeneracy of the ground state does not increase
exponentially, which is characteristic of the black hole. The reason
behind this is the fact that the collective modes of the black hole
are described by an effective string theory. The effective string
theory also incorporates small deviations from extremal and stable
black holes and can be used to study black hole thermodynamics for
very small values of the Hawking temperature.

The effective string theory picture of the collective modes of the
5-dim. black hole is most easily understood in the limit of weak
coupling when the D-brane assembly does not actually form a black
hole, because the Schwarzchild radius sits within the characteristic
length scale of the basic theory, namely the string length. However
ignoring the difficulties of the strongly coupled regime of the gauge
theory (which actually describes the black hole regime) one can do
various calculations. The most important set of calculations, besides
the ones that calculate the entropy of the excited black hole, set out
to compare the Hawking decay rates of various particles that interact
with the black hole.  In all cases the calculations done with the
effective string matched with those done using standard methods in
relativity. An important conceptual point here is that the answers of
black hole thermodynamics are obtained by using the standard rules of
quantum mechanics. One calculates the relevant S-matrix for the
interaction of a closed string with the effective string that
describes the collective modes of the black hole. Then the absolute
value square of the S-matrix is averaged over the initial micro-states
and the final micro-states states are summed over. These micro-states
are simply built out of the oscillators of the effective string.

Presently there is much interest in putting many of these calculations
on a firm footing, especially because their validity in the black hole
regime where $gN\sim 1$, is far from obvious.  There are indications
that the agreements at weak coupling may continue to hold at strong
coupling. This indicates that the D-brane bound state somehow knows
about the geometry of the black hole. To explicitely see the emergence
of the geometry of the black hole by constructing the effective theory
of a strongly interacting probe is a challenge for the future.

In our view, which is shared by many working in the subject, the
information paradox is likely to be resolved in the framework which we
have tried to give a glimpse of. In the process we have begun to
understand how the conventional ideas of space time and general
relativity have to be replaced at a fundamental level. They emerge at
scales larger than the string length, from a more fundamental
non-commutative geometry in which space time is described by
non-commuting matrices.

The unfolding story about the fundametal theory and its 
applications to black hole physics, that I have tried to sketch above, 
in this very concise account, is the work of a whole
community of people who were and are still called 'string
theorists'. Below I give a few references that may help navigate the
interested reader through the literature.

\vspace*{.4cm}

\noindent{\Large\bf{Acknowledgement}}

\vspace*{.3cm} 

I would like to thank Justin David, Avinash Dhar, S.Fawad Hassan and
Gautam Mandal for enjoyable discussions on the subject and Leena
Chandran for many helpful comments on the manuscript.

\end{document}